\documentclass[a4paper,fleqn,usenatbib]{mnras}

\usepackage[T1]{fontenc}
\usepackage{ae,aecompl}

\usepackage{graphicx}	
\usepackage{amsmath}	
\usepackage{amssymb}	

\newcommand{\be}{\begin{equation}}
\newcommand{\ee}{\end{equation}}
\newcommand{\ba}{\begin{eqnarray}}
\newcommand{\ea}{\end{eqnarray}}

\newcommand{\barr}{\begin{array}}
\newcommand{\earr}{\end{array}}

\newcommand{\refeqn}[1]{equation \eqref{#1}}

\title[GR Simulations of Transonic Accretion Flows]{General Relativistic Numerical Simulation of sub-Keplerian Transonic Accretion Flows onto Rotating Black Holes: Kerr Spacetime}

\author[J. Kim et al.]{
Jinho Kim,$^{1,2}$
Sudip K. Garain,$^{2}$\thanks{E-mail: sgarain@nd.edu}
Sandip K. Chakrabarti$^{3,4}$
and Dinshaw S. Balsara$^{2}$
\\
$^{1}$Korea Astronomy \& Space Science Institute, 776 Daedeokdae-ro, Yuseong-gu, Daejeon 34055, Korea \\
$^{2}$Department of Physics, University of Notre Dame, Notre Dame, IN 46556, USA\\
$^{3}$S. N. Bose National Center for Basic Sciences, Block JD, Sector III, Salt Lake, Kolkata, 700098, India\\
$^{4}$Indian Center for Space Physics, 43 Chalantika, Garia St. Rd., Kolkata, 700084, India
}

\date{Accepted XXX. Received YYY; in original form ZZZ}

\pubyear{2017}

\begin{document}
\label{firstpage}
\pagerange{\pageref{firstpage}--\pageref{lastpage}}
\maketitle

\begin{abstract}
We study time evolution of sub-Keplerian transonic accretion flows onto black holes using 
a general relativistic numerical simulation code. We perform simulations around the black holes
having non-zero rotation. We first compare one-dimensional simulation results with 
theoretical results and validate the performance of our code. Next, we present results of 
axisymmetric, two-dimensional simulation of advective flows.  In the literature,
there is no solution which describes steady shock solutions in two dimensions. However, our
simulations produce these centrifugal force supported steady shock waves even in presence 
of strong dragging of inertial frames. Since the post-shock region could be hot
and upscatter photons through Comptonization, these shock would put imprints on 
the spectra. Thus, our solutions, which represent truly new results, 
could be useful to measure spins through radiation spectrum of accreting 
Kerr black holes.
\end{abstract}




\section{Introduction}

Matter from a companion star in a binary system spirals towards the primary through a process known 
as accretion. In the case of high viscosity, matter injected at the outer boundary would form a 
Keplerian disk before entering into the black hole \citep{Chakrabarti1996a}. In the present paper, we study what happens 
when an inviscid flow of low angular momentum matter spirals into a rotating black hole, 
space-time geometry around which is the well known Kerr geometry. A low-angular momentum 
matter would feel the centrifugal barrier ($\sim l^2/r^3$, $l$ and $r$ being the specific 
angular momentum and the radial distance) and would slow down, and pile up eventually making a 
possible density jump, before entering the black hole. This is a possibility 
for inviscid flow, though in presence of viscosity the result depends on the exact
magnitude of viscosity. These shock waves, just outside the horizon 
provides an opportunity to study the emission of radiation in a strong gravity limit and would therefore be of great interest, especially to pinpoint the mass and the spin of the central object. 

In the two-component advective flow (TCAF) scenario \citep[][and references therein]{Chakrabarti1995a},
the soft photons from the
Keplerian disk are intercepted by the post-shock region and leave the system as hard radiation. The
oscillations of the post-shock region will leave its signature in the outgoing hard photons.
It has been shown using numerical simulations and theoretical work
\citep{Molteni1996a,Chakrabarti2004a,Garain2014a,Chakrabarti2015a} that when there is a resonance between the
infall time-scale (compressional heating) and the cooling time-scale, the shock surface oscillates at a particular
frequency which is approximately inverse to the infall (or, cooling) time-scale.
It was further shown in \citet{Garain2014a} that the resulting
light curves exhibit quasi-periodic oscillations (QPOs) and the frequency of the
QPOs are consistent with the infall time-scales \citep[see also][]{Chakrabarti2015a}.
Since the infall time-scale is proportional to the (radius)$^{3/2}$ of the shock, larger
shock radius produces lower QPO frequency. In fact, TCAF
model has been added as a local additive table model into HEASARC`s spectral
analysis software package XSPEC.
Consequently, for outbursting sources where QPO frequency starts from mHz
and continue to increase to a few 10s of Hz, it has been found that the shock forms far away
from the black hole initially and then propagates towards to black hole during the rising phase
\citep{Debnath2014a,Chatterjee2016a,Debnath2017a,Bhattacharjee2017a}.
In the declining phase, the opposite phenomena occurs, namely, the shock recedes and the QPO frequency steadily goes down.
Several physical parameters associated with the accretion disks such as accretion rates, Mass of the black holes,
shock locations and the strength of the shock are found by analyzing the data for various sources (see above refs).

Global solutions of weakly viscous transonic flows in Kerr geometry have been been found in great detail by \citet{Chakrabarti1996c,Chakrabarti1996b}. Complete theoretical studies for accretion and winds are obtained assuming the flow to be in vertical equilibrium (VE) or in the form of conical wedge (CF). 
It was found that for a large region of the parameter space in each model, 
a shock wave will form as a distance depending on the conserved energy, angular momentum and spin parameters.
These solutions give the closest possible description to an actual flow dynamics, since a low angular momentum 
flow is likely to be geometrically thick and quasi-spherical far away from the black hole. 
However, theoretical works cannot be done in such cases.

In this paper, we concentrate on the numerical simulations of these flow in a complete Kerr geometry. 
Our goal is first to see if the theoretical results mentioned above are instead stable. This is because of the fact that a flow actually slows down to a large extent in the post-shock region, just outside the horizon, itself is very exciting and it would be important to through light on their stability properties. 
Second, we would also like to prove the effects of the dragging of inertial frames and eventually, whether such effects cause precession of outflows and jets. So the code must be tested against known theoretical solutions. We will test the code for a conical flow for simplicity. Since the relevant equations are given in \citet{Chakrabarti1996c,Chakrabarti1996b}, we do not repeat them here. Instead we only mention the key points. 

In this paper, we choose $R_g=GM_{\rm{BH}}/c^2$ as the unit of length, $R_g c$ as unit of angular momentum, and $R_g/c$ as unit of time. In addition, we choose the geometric units $G=M_{\rm{BH}}=c=1$ ($G$ is gravitational constant, $M_{\rm{BH}}$ is the mass of the central black hole and $c$ is the unit of light). Thus $R_g=1$, and angular momentum and time are measured in dimensionless units. 

\section{Analytical Solution}
\label{sec:sec1}

We assume a thin conical wedge shaped adiabatic flow symmetrically placed on 
both sides of the equatorial plane, $\theta=\pi/2$, entering into a Kerr black hole.
For analytical studies,  we assume the Kerr metric transformed to the 
cylindrical coordinate which is written as follows \citep{Novikov1973a},
\begin{equation}
\label{eqgmunu}
\begin{split}
ds^2&= g_{\mu\nu}dx^{\mu}dx^{\nu}\\
&= - \frac{r^2\Delta}{A}dt^2 + \frac{A}{r^2}\left( d\phi - \omega dt\right)^2
+ \frac{r^2}{\Delta}dr^2 + dz^2.
\end{split}
\end{equation}
Here,
$$
A=r^4+r^2a^2+2ra^2,~\Delta=r^2-2r+a^2,~\omega=2ar/A, 
$$
$a$ being the spin parameter of the black hole and $\omega$ represents the frame dragging due to the rotation of the central black hole.

In absence of viscosity and any heating or cooling, one can find the 
conserved specific energy as \citep{Chakrabarti1996b} 
\begin{equation}
\label{eqE}
\epsilon =hu_t=\frac{1}{1-na_s^2}u_t,
\end{equation}
where, $n=1/\left(\Gamma-1\right)$ is the polytropic index, $\Gamma$ being the adiabatic index
and  $h=1/\left(1-na_s^2\right)$ is the enthalpy, $a_s$ being the sound speed. Also,
\begin{equation}
u_t=\left[\frac{\Delta}{(1-V^2)(1-\Omega l)(g_{\phi\phi}+l g_{t\phi})}\right]^{1/2}.
\end{equation}
Here,
\begin{equation}
\label{omega}
\Omega = \frac{u^\phi}{u^t}=-\frac{g_{t\phi}+lg_{tt}}{g_{\phi\phi}+lg_{t\phi}},
\end{equation} 
and $l=-u_\phi/u_t$ is the specific angular momentum. 
Also, 
\begin{equation}
\label{bigV}
V=\frac{\mathcal{V}}{\left(1-\Omega l\right)^{1/2}},
\end{equation} 
where 
\begin{equation}
\label{smallv}
\mathcal{V}=\left(-\frac{u_ru^r}{u_tu^t}\right)^{1/2}.
\end{equation}

We rewrite equation (\ref{eqE}) as
\begin{equation}
\label{eqE1}
\epsilon =\frac{1}{1-na_s^2}\frac{1}{\left(1-V^2\right)^{1/2}}F(r),
\end{equation}
where, 
\be
F(r) =\left[\frac{\Delta}{(1-\Omega l)(g_{\phi\phi}+l g_{t\phi})}\right]^{1/2}.\nonumber
\ee

The entropy accretion rate \citep{Chakrabarti1989a,Chakrabarti1996b} can be obtained as
\begin{equation}
\label{eqM}
\dot{\mu}=\left(\frac{a_s^2}{1-na_s^2}\right)^n \frac{V}{\left(1-V^2\right)^{1/2}}G(r),
\end{equation}
where, $G(r) =r\Delta^{1/2}$.

We follow the standard solution procedures \citep{Chakrabarti1996c,Chakrabarti1996b} to calculate
$V(r)$ and radial dependence of other required quantities. By differentiating equations (\ref{eqE1}) 
and (\ref{eqM}) with respect to $r$ and eliminating terms involving $da_s/dr$, we find following
expression as the gradient of $V(r)$:
\begin{equation}
\label{eqdVdr}
\frac{dV}{dr}=\frac{V\left(1-V^2\right)\left[a_s^2R_1-R_2\right]}{\left(V^2-a_s^2\right)},
\end{equation}
where, $R_1=\frac{1}{G(r)}\frac{dG(r)}{dr}$ and $R_2=\frac{1}{F(r)}\frac{dF(r)}{dr}$.

At the sonic point, both numerator and the denominator vanish and one obtains 
the so-called sonic point condition as
\begin{equation}
\label{soniceqn}
V_c= a_{s,c}\,;  \quad
a_{s,c}^2 = \left. \frac{R_2}{R_1}\right|_c.
\end{equation}
Here, subscript $c$ refers to the quantities evaluated at the sonic point $r=r_c$.

To find a complete solution from the horizon to infinity for a given black hole spin
parameter $a$, one needs to supply the
specific energy $\epsilon$ and the specific angular momentum $l$.
For Kerr black holes, a few examples of classification of parameter space spanned 
by these two parameters, namely $\epsilon$ and $l$, has been provided in \citet{Chakrabarti1996c,Chakrabarti1996b}.
The accretion or wind solutions may pass through one or multiple sonic points.
Moreover, a solution having more than one sonic points may form a standing shock.
The location of the shock may be found by solving continuity of energy equation,
mass flux equation and relativistic momentum balance condition simultaneously. These are
collectively called the `shock conditions'. 
For more discussions on these solutions, readers are referred to the 
aforementioned references. In this paper, we obtain shocks locations in some flow parameters and
compare the results of numerical simulations to see if the shocks indeed form in a 
realistic flow.

\section{Numerical Simulation Procedure}
\label{sec:sec2}
In this section, we describe how we can obtain the time dependent numerical solution using the boundary condition provided in \autoref{sec:sec1}.
Instead of the spacetime metric provided in \refeqn{eqgmunu} we use the usual Boyer-Lindquist coordinates $\left(t,r,\theta,\phi\right)$ for the Kerr metric  for the numerical simulation \citep{Boyer1967a}. 
In the ADM 3+1 decomposition, the spacetime is expressed in terms of the lapse ($\alpha$), shift ($\beta^i$) and spatial metric ($\gamma_{ij}$) \citep{Arnowitt2008a}.
The line element around Kerr black hole can be expressed as follows:
\be
\label{eq:metric}
\begin{split}
ds^2 =& -\alpha^2 dt^2 + \gamma_{ij}\left(dx^i+\beta^i dt \right)\left(dx^j+\beta^j dt \right)\\
=& -\left(1-\frac{2r}{\rho^2}\right)dt^2 -\frac{4ar\sin^2\theta}{\rho^2}dt d\phi \\
 &  + \frac{\rho^2}{\Delta}dr^2 + \rho^2 d\theta^2 + \frac{\Sigma}{\rho^2} \sin^2\theta d\phi^2,
\end{split}
\ee
where 
\be
\begin{split}
\rho^2=&r^2+a^2\cos^2\theta, \\
\Delta=&r^2-2r+a^2. \\
\Sigma=&\left(r^2+a^2\right)^2-a^2\Delta\sin^2\theta. \nonumber
\end{split}
\ee
Then, the lapse, $\alpha$, and shift, $\beta^i$, in equation \refeqn{eq:metric} are
\be
\begin{split}
&\alpha=\sqrt{\frac{\rho^2 \Delta}{\Sigma}}, \\
&\beta^r = \beta^\theta=0, \quad \beta^\phi = -\frac{2ar}{\Sigma}.
\end{split}
\ee

In this paper, we do not consider the evolution of the spacetime metric by the influence of the accreting matter.
In other words, we assume that the accretion mass is too light to affect the the spacetime metric of the central black hole.
For typical accretion disc around a central black hole of Mass 10 $M_\odot$, the mass accretion rate ($\dot{M}$) of one Eddington rate, 
when expressed in the unit of $M_\odot$, comes out to be $\sim10^{-15}M_\odot/$s.
Such a small accretion rate cannot increase the mass or change the angular momentum of the central black hole 
within our simulation time ($\sim$ a few 10 milliseconds).
Therefore, we can rightly neglect the self gravity of the disc.

For the evolution of the accreting matter (hydrodynamic simulation), we adopt the Valencia formulation which is written as a flux conservative form in the 3+1 decomposition of spacetime \citep{Banyuls1997a}.
In our coordinate system, the conservative ($q$) and primitive ($w$) variables are,
\be
q=\left(
\begin{array}{c}
D \\ S_r \\ S_{\theta} \\ S_{\phi} \\ \tau
\end{array}
\right)
=\left(
\begin{array}{c}
\rho_0 W \\ \rho_0 h W^2 v_r \\ \rho_0 h W^2 v_{\theta} \\ \rho_0 h W^2 v_{\phi} \\ \rho_0 h W^2 -P -D
\end{array}
\right),\,
w=\left(
\begin{array}{c}
\rho_0 \\ v^r \\ v^{\theta} \\ v^{\phi} \\ P
\end{array}
\right).
\label{eq11}
\ee
Here, $\rho_0$ is the fluid rest mass density, $P$ is the pressure and $h$ is the specific enthalpy. 
They are measured in the co-moving frame of the fluid.
$v^i$ is the fluid velocity measured by Eulerian reference frame i.e., $v^i = \frac{u^i}{\alpha u^t}+\frac{\beta^i}{\alpha}$.
$W$ is the Lorentz factor and defined as $W=\alpha u^t=1/\sqrt{1-\gamma_{ij} v^i v^j}$.
Here, $\gamma_{ij}$ are the spatial part of the metric components $g_{\mu\nu}$.
The radial and angular velocity in the Eulerian frame can be expressed in 
terms of $v$ in \refeqn{smallv} and $\Omega$ in \refeqn{omega}:
\be\label{vrvphi}
\begin{split}
v^r&=\sqrt{-\frac{g_{tt}+g_{t\phi}\Omega}{g_{rr} \alpha^2} }v \\
v^{\phi}&=\frac{1}{\alpha}\left(\Omega-\beta^\phi \right)
\end{split}
\ee

Assuming axisymmetry ($\frac{\partial}{\partial\phi}=0$), the hydrodynamic equations in 
the curved spacetime that described in \refeqn{eq:metric} can be written as follows:
\be
\frac{\partial\left(\sqrt{\gamma}q\right)}{\partial t}
+\frac{\partial\left(\sqrt{-g}f^r\right)}{\partial r}
+\frac{\partial\left(\sqrt{-g}f^{\theta}\right)}{\partial \theta}
=\sqrt{-g}S,
\label{eq:hydro1}
\ee
where
\ba
f^r&=&\left[
\begin{array}{ccccc}
Dv^r\\S_rv^r+P\\S_{\theta}v^r\\S_{\phi}v^r\\\tau v^r+P v^r
\end{array}
\right],\nonumber\\
f^{\theta}&=&\left[
\begin{array}{ccccc}
Dv^{\theta}\\S_rv^{\theta}\\S_{\theta}v^{\theta}+P\\S_{\phi}v^{\theta}\\\tau v^{\theta}+Pv^{\theta}
\end{array}
\right],\nonumber\\
S&=&\left[
\begin{array}{ccccc}
0\\
\frac{1}{2}T^{\mu\nu}\partial_r g_{\mu\nu}\\
\frac{1}{2}T^{\mu\nu}\partial_\theta g_{\mu\nu}\\
0\\
\alpha\left(T^{\mu t}\partial_\mu(\ln\alpha)-T^{\mu\nu}\Gamma^t_{\mu\nu}\right)
\end{array}
\right]. \label{eq:hydro2}
\ea
Here $T^{\mu\nu}$ is the stress energy tensor of the perfect fluid which is defined as $T^{\mu\nu}=\rho_0 h u^{\mu}u^{\nu}+Pg^{\mu\nu}$.
$\sqrt{\gamma}$ and $\sqrt{-g}$ are the determinants of spatial and spacetime metric, respectively.
From the Kerr metric shown in \refeqn{eqgmunu}, we have 
\be
\sqrt{\gamma}=\sqrt{\frac{\rho^2 \Sigma}{\Delta}}\sin\theta,\quad
\sqrt{-g}=\alpha\sqrt{\gamma}=\rho^2\sin\theta.
\ee
As we see in \refeqn{eq:hydro2}, this formulation is not strictly a flux conservative form due to the non-zero source terms in $S$. 
The source terms, however, contain spatial derivatives of the metric components only (they come from the Christoffel symbols)
and they do not have any derivatives of the hydrodynamic variables.
Note that even for the general time dependent metric, the time derivative of metric components can be substituted 
by their spatial derivatives from the Einstein equation. Therefore, the source term does not have time derivatives.

We use the ideal gas equation of state which can be written in the following form:
\be\label{eq30}
P=\left(\Gamma-1\right)\rho_0 e,
\ee
where $e$ is the specific internal energy.
The above equation of state provides the expression of specific enthalpy:
\be\label{findP}
h=1+\frac{\Gamma}{\Gamma-1}\frac{P}{\rho_0}.
\ee 

In this paper, we solve the hydrodynamic equations in (\ref{eq:hydro1}) and (\ref{eq:hydro2}) using a numerical code developed by \citet{Kim2012a}.
The details of the code can be found in \citet{Kim2012a}.
The most useful property of this code is that it can be applied to any spacetime 
metric in any coordinate system. 
Although the original code by \citet{Kim2012a} is written in the so-called ``pseudo-Newtonian'' metric, we can easily apply the code to the Kerr geometry.
This code uses finite volume method to enforce local conservation of the fluid in the computational grid.
Therefore, the code can guarantee total mass and angular momentum conservations which appear in the first and fourth rows of \refeqn{eq:hydro2}.
For the treatment of the discontinuous solution of the fluid such as shocks, rarefactions or contact discontinuities, the High Resolution Shock Capturing (HRSC) techniques are applied in the code.
We use the third order slope limiter proposed by \citet{Shibata2003a} which is based 
on the {\it minmod} function. For the flux approximation, we use the HLL method that takes care of maximum and minimum wave speed \citep{Harten1983a}. 
The HLL method has some dissipation but the results are very stable.
For the time integration, we use the third order three stage Strong Stability-Preserving (SSP) 
Runge-Kutta method by using method of line. It is known as Shu-Osher method \citep{Shu1988a}.

\section {Results}
\label{sec:sec3}

\subsection{One dimensional conical accretion flow}

First we test the code by reproducing theoretically obtained steady non-linear solutions using the code. We test for both the accretion and winds. In the first case, the accreting matter starts from a large 
distance and after passing through a sonic point, it passes through the 
standing shock where many of the flow variables jump abruptly. After that, it passes through another
mandatory sonic point before disappearing behind the horizon supersonically. In the second case, the wind
starts from an accretion flow and due to thermal pressure, it is pushed out to infinity after successively passing through the innermost sonic point, a shock, and the outer sonic point. The challenges faced by a code are to
test if (i) the shocks are actually formed, and if yes, (ii) whether the
jump in Mach number occurs in a very narrow region and exactly at the same place as predicted, since theoretically, the
shocks in an inviscid flow are infinitesimally thin. Once the tests are successful, one could explore
unknown territories, such as geometrically thick, fully two dimensional flows.

\begin{figure*}
\includegraphics[width=\columnwidth]{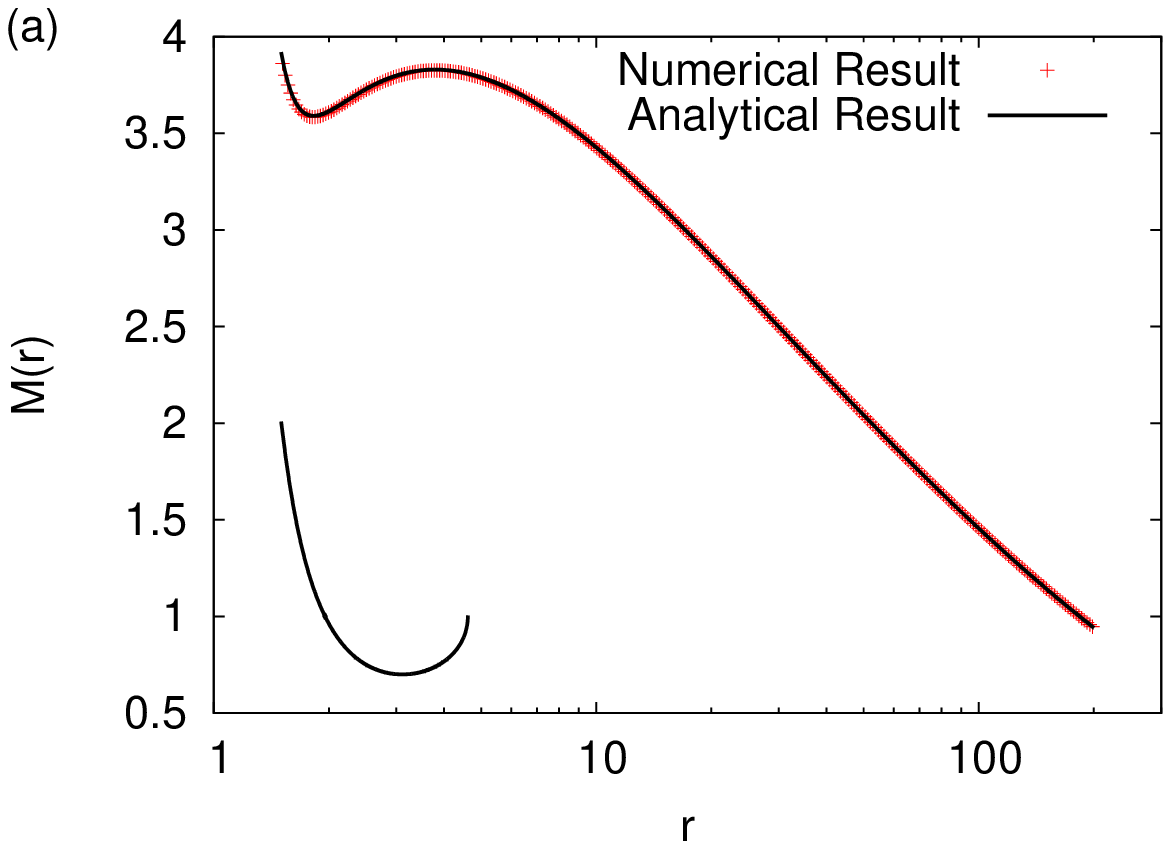}
\includegraphics[width=\columnwidth]{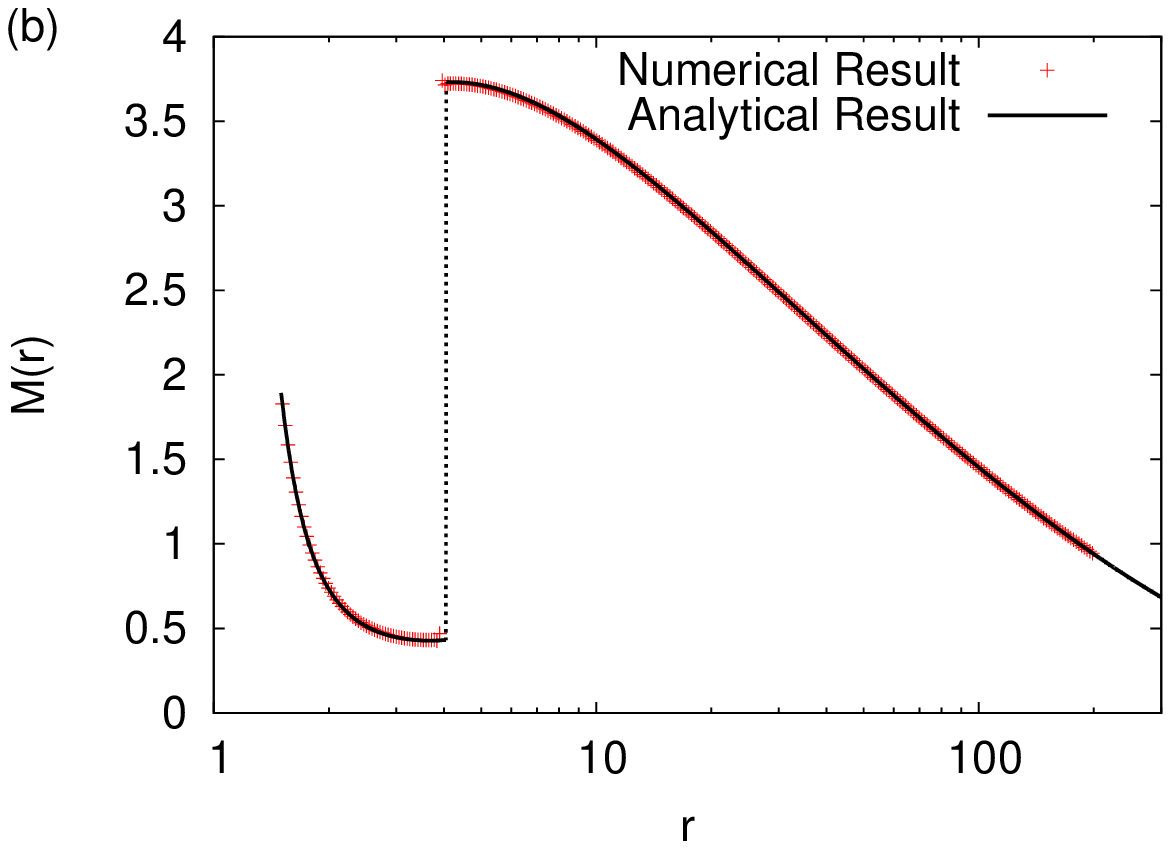}
\includegraphics[width=\columnwidth]{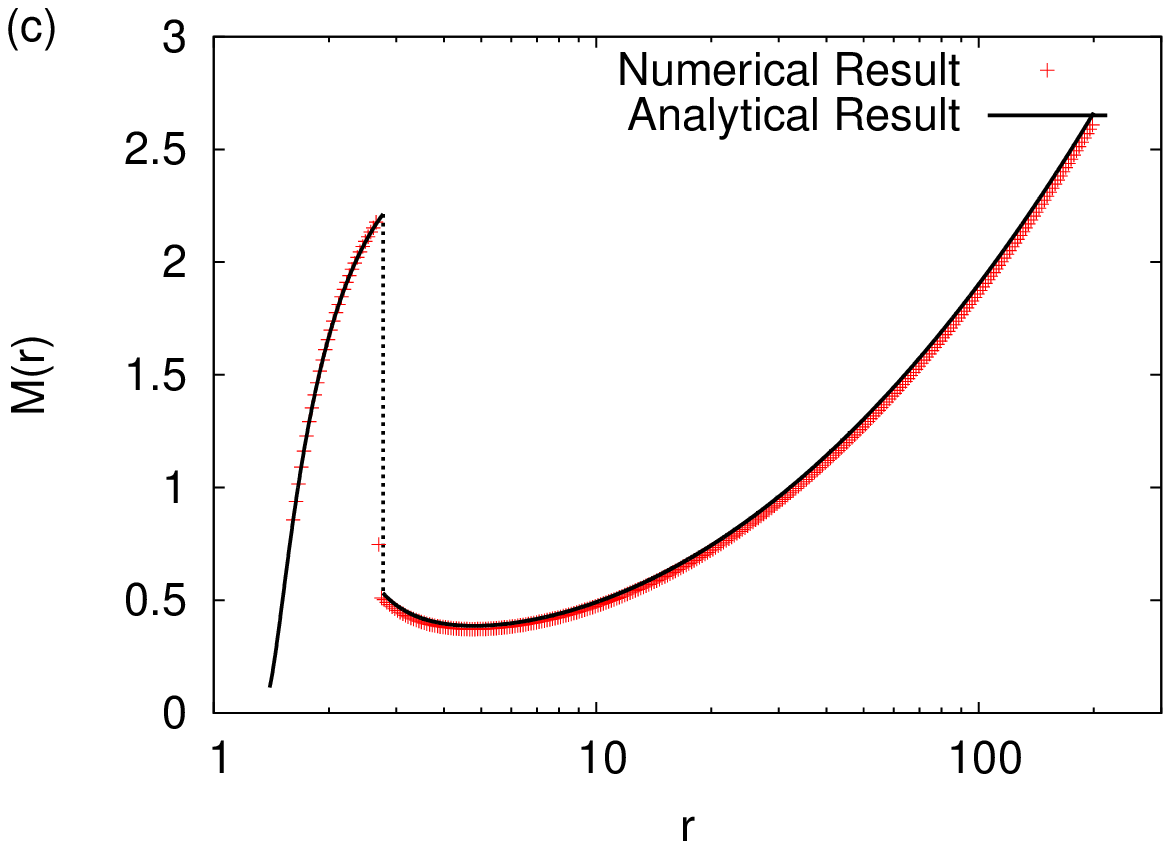}
\includegraphics[width=\columnwidth]{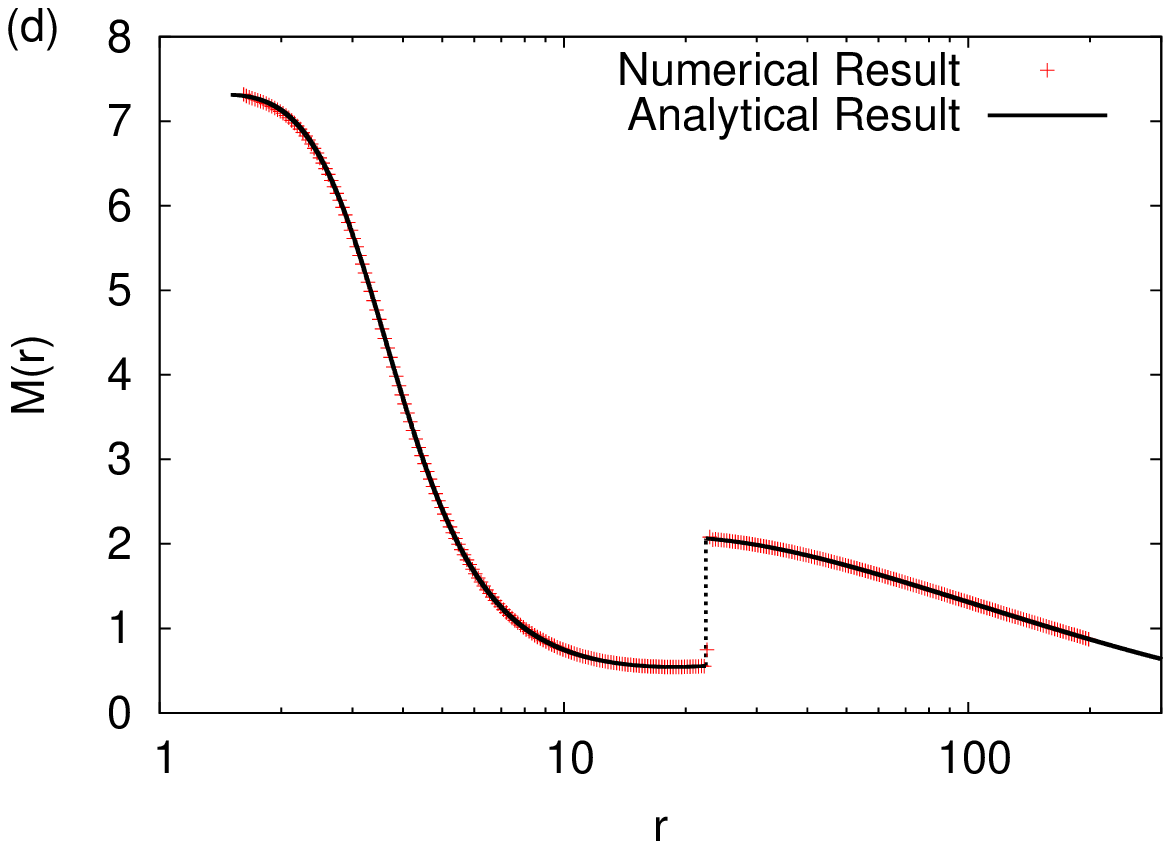}
\caption{Comparison of radial Mach number ($V/a_s$) variation for one dimensional axisymmetric accretion and wind 
solutions around a rotating black hole. Black solid lines are obtained using analytical method, 
whereas red points are the simulation results.
In (a), we present an accretion solution without a shock. $\epsilon=1.004$ and $l=2.2$ are chosen 
for this solution (NSA region). (b) An example of an accretion solution
where a shock is formed. $\epsilon=1.004$ and $l=2.25$ are chosen for this solution (SA region).
According to analytical calculation, shock location is found to be at $r=4.05$. 
(c) An example of the wind solution where a shock is formed at $r=2.77$. 
The parameters for this solution are $\epsilon=1.02$ and $l=2.3$ (SW region of parameter).
The spin parameter $a=0.95$ for all three cases. 
(d) An example of an accretion solution where a shock is formed for a retrograde flow.
$\epsilon=1.004$, $l=4.0$ and spin parameter $a=-0.95$ are used for this case. Analytically, the shock location 
is found to be at $r=22.45$. Numerical simulations clearly captured the shocks exactly where theoretical 
shock location was predicted.
}
\label{fig:1d}
\end{figure*}

In \autoref{fig:1d}(a)-(c) we show the results of three numerical simulations for the prograde flows. 
There were 300 logarithmically spaced radial grids for all the one dimensional simulations presented here.
For accretion solutions, the outer
and inner boundaries are located at $r_{\rm out}=200$ and $r_{\rm in}=1.2$, respectively.
We use $a=0.95$ for all three cases. In \autoref{fig:1d}(a), we consider the flow parameter 
from the ``No Shock in Accretion'' (NSA) region \citep[Figure 2 of][]{Chakrabarti1996b}. 
If a flow starts with parameters from this region, then the shock conditions are not 
satisfied and the flow passes through the outer sonic point 
and enters into the black hole smoothly. This particular case is with the parameter pair 
of $\epsilon=1.004$ and $l=2.2$. The solid black curves give the theoretical distribution of the 
Mach number ($V/a_s$) as a function of the radial distance. The upper branch is complete in the sense that 
it connects the black hole horizon with a large distance, while the lower (subsonic) branch is 
an isolated solution which does not connect flows at a large distance. Red `+' signs signify the 
results of numerical simulations. Matter is injected at $r_{\rm out}=200$ with the radial velocity 
and sound speed as in the theoretical solution. We find that the steady solution exactly 
matched with the theoretical solution branch. 

In the next simulation (\autoref{fig:1d}(b)), we choose the injected parameter from a region 
called `Shock in Accretion' (SA) of \citet{Chakrabarti1996b}. We choose $\epsilon=1.004$ and 
$l=2.25$ in this case. The analytical solution is exactly reproduced by the numerical simulation 
with the shock thickness about a grid size which is
the smallest resolvable length scale. The shock is formed at $r=4.05$. The reason why the flow chose to jump on the
lower branch is that the entropy of the flow which passes through the inner sonic point 
is higher than that of the upper branch and flow chose that branch for stability. 
In \autoref{fig:1d}(c), we carry out a simulation with flow parameters from the `Shock in Winds' (SW) region. 
Here the parameters are $\epsilon=1.02$ and $l=2.3$. The matter is injected from a
radius $r_{\rm in}=1.5$ outwards on the equatorial plane assuming there is a matter source (such as a
disk). It immediately passes through the sonic point, but reaches to a large distance only after 
passing through a shock wave. When comparing with the theoretical solution, we note that there is a slight 
deviation at a large distance. On inspection, we find that this is due to finite resolution of the grid. A small
error (which is equivalent to a large fractional error) in the location of shock wave 
(which is grid resolution limited) is propagated downstream and affect the numerical results. 
The error in thermodynamic quantities in the shock location does not propagate backward in the 
supersonic region, and thus does not affect the pre-shock branch at all. The shock is located at 2.77.
Clearly the error would be reduced with the reduction of the grid size.

In \autoref{fig:1d}(d), we present the results of our simulation for a retrograde accretion flow.
We choose $\epsilon=1.004$, $l=4.0$ and spin parameter $a=-0.95$ in this case. The outer
and inner boundaries are located at $r_{\rm out}=200$ and $r_{\rm in}=1.6$, respectively.
Analytical method predicts the shock to form at $r=22.45$. As in the prograde flow, here also, we
find the analytical solution is exactly reproduced by the numerical simulation with the shock thickness 
about a grid size.

\subsection{Two dimensional Bondi accretion flow}

Having convinced ourselves that the numerical code has reproduced theoretical nonlinear solution 
exactly, we turn our attention to solve the simplest possible two dimensional problem which does not
have any theoretical solution. For all the two dimensional flows presented here, 
there were 300 logarithmically spaced radial grids and 90 equal spaced grids along the polar angle. 
We study the spherically symmetric flow (at a large distance) on a Kerr geometry
which is a special case of \citet{Chakrabarti1996b} 
solution for $l=0$. Earlier \citet{michel1972} 
solved for Bondi
flows on a Schwarzschild geometry which remained spherically symmetric even on the horizon. However,
as we show below, due to the dragging of inertial frame the flow becomes necessarily axisymmetric rather 
than spherically symmetric close to the black hole.
Our goal would be to see how the dragging of the inertial frame would modify the symmetry of the flow. 
 
To show the effects of dragging of frames due to spin of the black hole, we run the same case twice: For the Kerr 
($a=0.95$) and the Schwarzschild ($a=0$) geometries and then take the difference in density distribution in the two cases.
We injected matter with radial velocity 0.02288 and sound speed 0.08053 at $r_{\rm out}=200$ \citep{Kim2017a}.
However, for Schwarzschild black hole case, the inner boundary is placed at $r_{\rm in}=2.1$ and for Kerr black hole,
$r_{\rm in}=1.5$.

\begin{figure}
\includegraphics[width=\columnwidth]{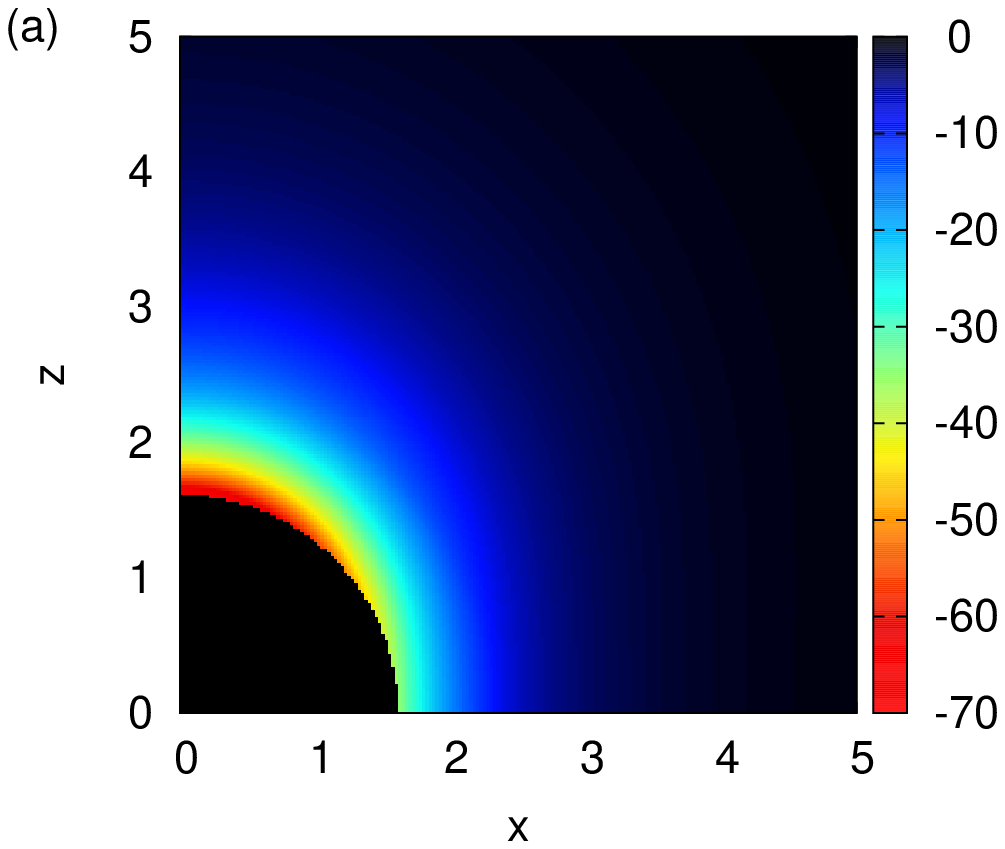}
\includegraphics[width=\columnwidth]{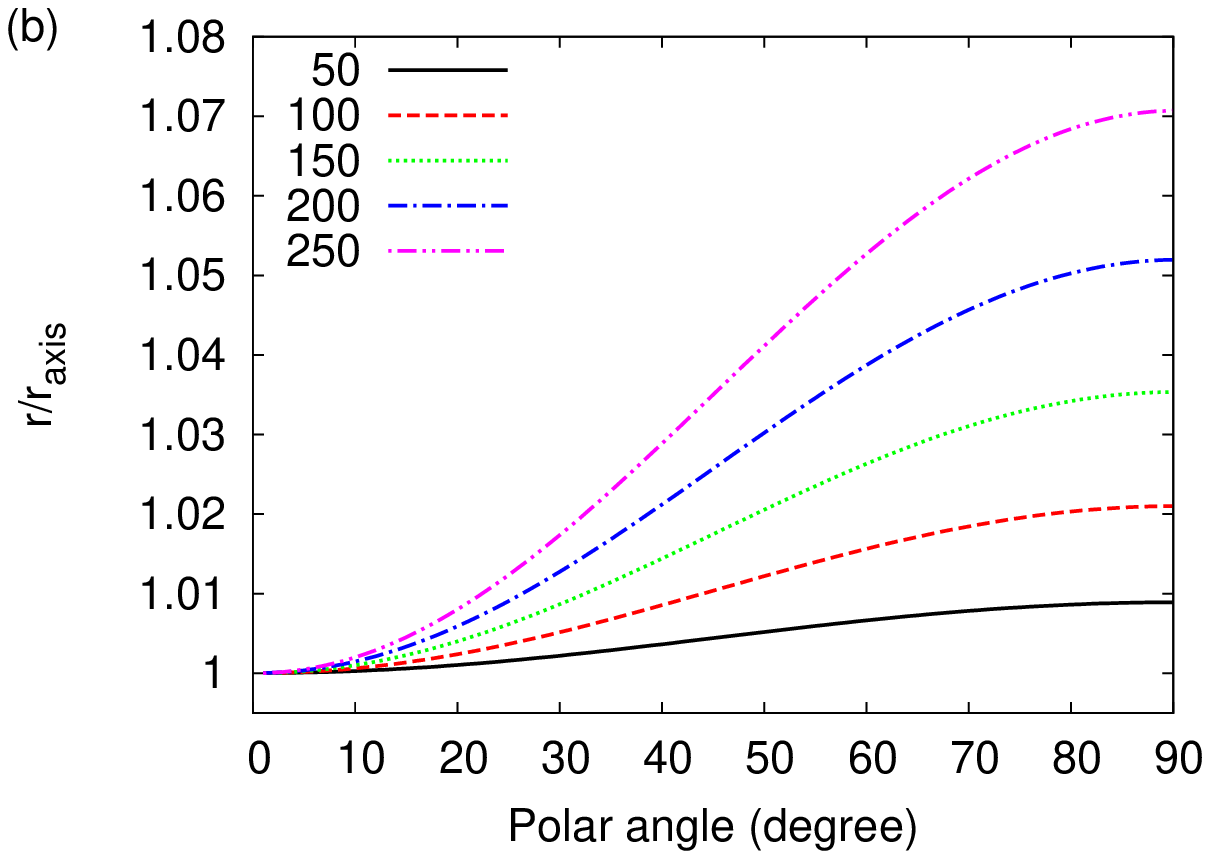}
\caption{Breaking of symmetry of the flow due to dragging of inertial frame is demonstrated here. (a) 
Distribution of difference in flow density with identical initial condition
at the outer grid at $r=200$ between a Kerr ($a=0.95$) black hole and a Schwarzschild ($a=0$) black hole
even if matter is accreted spherically symmetrically far from the central object. 
We note that while near the equatorial plane the difference is very little, highest difference occurs 
on the pole. In (b) we plot the contours of constant density in the simulation result as a function of the 
polar angle and the ratio of the distance (of a given density) divided by the 
distance (of the same density) along the pole. 
We see that the same high density $250$ occurs on the pole at about 7\% distance less than the distance on the equatorial plane.
The contours are, from bottom to top,  $50$, $100$, $150$, $200$ and $250$ respectively.}
\label{zerol2d}
\end{figure}

In \autoref{zerol2d}(a), we plot colours of the density difference. We see that the difference is higher along the axis. 
This is expected since the effects of spin on space time is the highest at the pole. 
In \autoref{zerol2d}(b), we plot the ratio of the radial distance for a given density divided by that on the polar axis
as a function of the polar angle. 
The contours of density $50$, $100$, $150$, $200$ and $250$ are shown (from bottom to top).
Clearly, higher the density, i.e., closer the flow to the horizon, 
the difference between the equatorial value and axial values are higher. Dragging of the
inertial frame induces a little rotation ($\Omega\sim 2.375\times 10^{-7}$) at the outer boundary of the grid on the equatorial 
plane where $r=200$.

\subsection{Two dimensional accretion flow with non-zero angular momentum}

\begin{table}
	\centering
	\caption{Parameters used for the two dimensional simulations.}
	\label{table:1}
	\begin{tabular}{ccccc} 
\hline Case & $\epsilon$, $l$ & $a$ & $V$ & $a_s$ \\
\hline
R1 & 1.001, 2.25 & 0.95 & 0.0634 & 0.0362 \\
R2 & 1.001, 4.0 & -0.95 & 0.0609 & 0.0363 \\
R3 & 1.001, 2.21 & 0.95 & 0.0634 & 0.0362 \\
R4 & 1.001, 2.63 & 0.95 & 0.06299 & 0.0362 \\
\hline
	\end{tabular}
\end{table}
We now apply the code to a more complex case where the accreting matter has some angular momentum. Such situations occur in
wind-fed systems as in Cygnus X-1. In the context of pseudo-Newtonian geometry, \citet{Molteni1994a} 
first carried out this simulation using Smoothed Particle Hydrodynamics (SPH) code and found that 
a standing shock is formed with an outflow along the polar direction. However, in the context of a complete Kerr geometry
the code has never been tested which formed non-linear shocks right outside a black hole horizon.  
We take $a=0.95$ black hole which means the horizon is located at $r=1.31$. We inject matter at the outer edge at $r=200$ and 
the inner absorbing boundary is chosen at $r_{\rm{in}}=1.5$. 
We choose 300 logarithmically spaced radial grid and 90 equispaced polar grids as before. 
In \autoref{table:1}, we present the parameters for all the two-dimensional runs presented here.
For run R1, we inject the flow with specific energy and angular momentum  $\epsilon=1.001$
and $l=2.25$ respectively. This pair of parameters produces a standing shock when vertical 
equilibrium model is used \citep{Chakrabarti1996c}.
At the outer radius, the injection velocity and sound speed are $V=0.0634$ and $a_s=0.0362$ in this case. We run for 
a time of $2.6\times10^4$, which is about ten times of the in-fall time.

\begin{figure}
\includegraphics[width=\columnwidth]{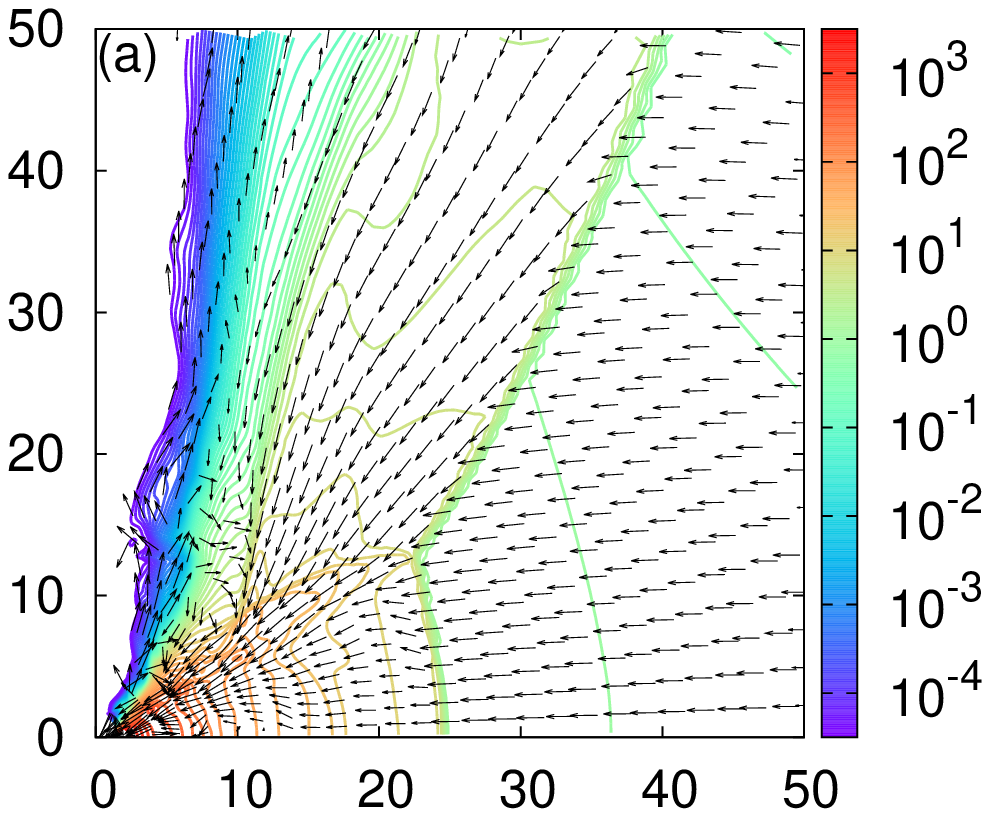}
\includegraphics[width=\columnwidth]{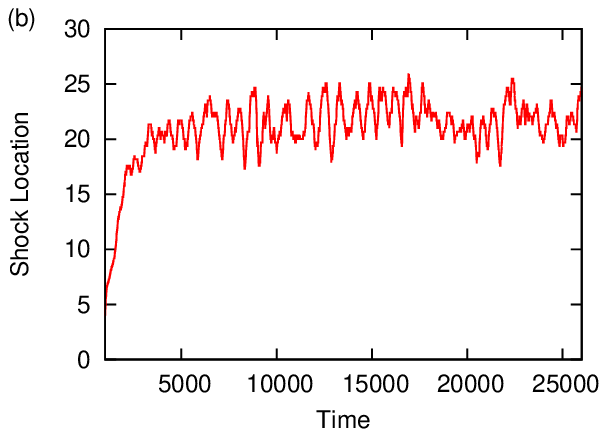}
\caption{ (a) Zoomed snapshot of density contours over-plotted
with velocity vectors for two-dimensional flow around a Kerr ($a=0.95$) black hole.
The parameters corresponding to run R1 are used.
The length of the arrow is proportional to the logarithm of the velocity magnitude.
(b) Time variation of shock location on the equatorial plane. A small amplitude oscillation
is see around a mean location $\sim 22$. See text for details.}
\label{fig:SA}
\end{figure}

We now show the results, including the initial transient phase when the matter rushed towards the black hole.  
In \autoref{fig:SA}(a), we show the inner region of the computational grid to show details of what happened after quasi-steady state
is achieved. We plot density contours superposed with velocity vectors. The length of the arrow is proportional 
to the logarithm of the velocity magnitude. A thin flow in a vertical equilibrium, 
a shock would be expected at $r=7.5$. However, we are using a geometrically 
thick flow which after passing through the oblique shock off the equatorial plane, 
converges towards the black hole and hits the centrifugal barrier 
and bounce back which then interacts with matter in-flowing on the equatorial plane. This turbulent pressure
along with the excess heat generated due to the fall of matter from off the equatorial plane pushes the shock outwards
to a distance $\sim 22$. \autoref{fig:SA} (b) shows the variation of the shock location on the equatorial plane. There is a 
small amplitude oscillation mostly driven by the inequality between the sum of the pressures in the pre- and 
post-shock regions. Note the formation of a strong outflow from the surface of the in-flowing post-shock region. 
Due to dominance of the centrifugal force, this region is also called the CENtrifugal pressure Supported BOundary Layer,
or CENBOL, which truly acts like a boundary layer of stars from which winds originate
and which up-scatters seed photons to higher energy \citep{Chakrabarti1995a}.

For this case, we can estimate the velocity of the outflowing matter at $r_{\rm{out}}$
and compare it with the escape velocity of at that radius.
The escape velocity at the outer computational domain is
\begin{equation}
v_{\rm{esc}} = \sqrt{\frac{2}{r}}.
\end{equation}
Accordingly, the escape velocity at the computational boundary ($r_{\rm out}=200$) is $v_{\rm{esc}} = 0.1$.
The fluid velocity is given by,
\begin{equation}
v = \sqrt{\gamma_{ij}v^iv^j}.
\end{equation}
\begin{figure}
\begin{centering}
\includegraphics[width=\columnwidth]{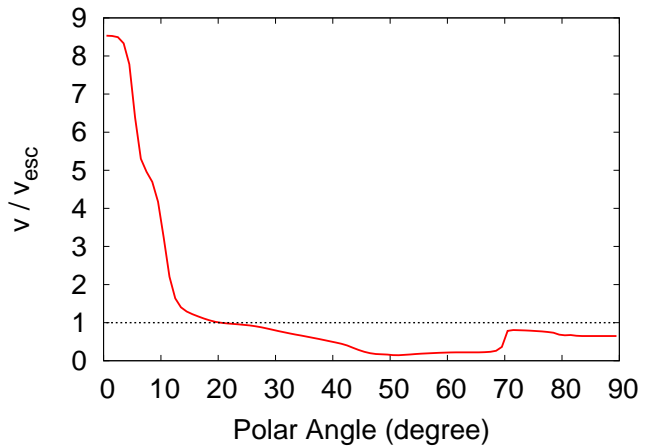}
\caption{Fluid velocity as a function of polar angle at the computational 
outer boundary ($r_{\rm out}=200$). The dotted horizontal line represents 
the escape velocity at $r_{\rm out}=200$.}\label{fig:vesc}
\end{centering}
\end{figure}
Figure \ref{fig:vesc} shows the fluid velocity at the outer boundary as a 
function of the polar angle for the case R1. The horizontal line shows the escape velocity for the comparison.
The numbers show that the fluid velocity is higher than the escape velocity at some locations close to the pole.
Near the pole, the fluid actually escapes from the vicinity of the black hole by forming a jet whereas the fluid at 
all other parts may fall back and become part of the accreting matter.
In our case, the outflow is launched from the post-shock region by the combined 
effects of thermal pressure and centrifugal pressure. As the flow passes through
the shock, the matter becomes hotter and entropy rises. This hotter, rotating flow is further
squeezed into a small volume due to compression as it further moves towards the black hole.
High entropy forces the flow to have the inner sonic point very close to the black hole,
where it becomes supersonic.
It subsequently expands and expelled in the vertical direction as a strong outflow.

\begin{figure}
\includegraphics[width=\columnwidth]{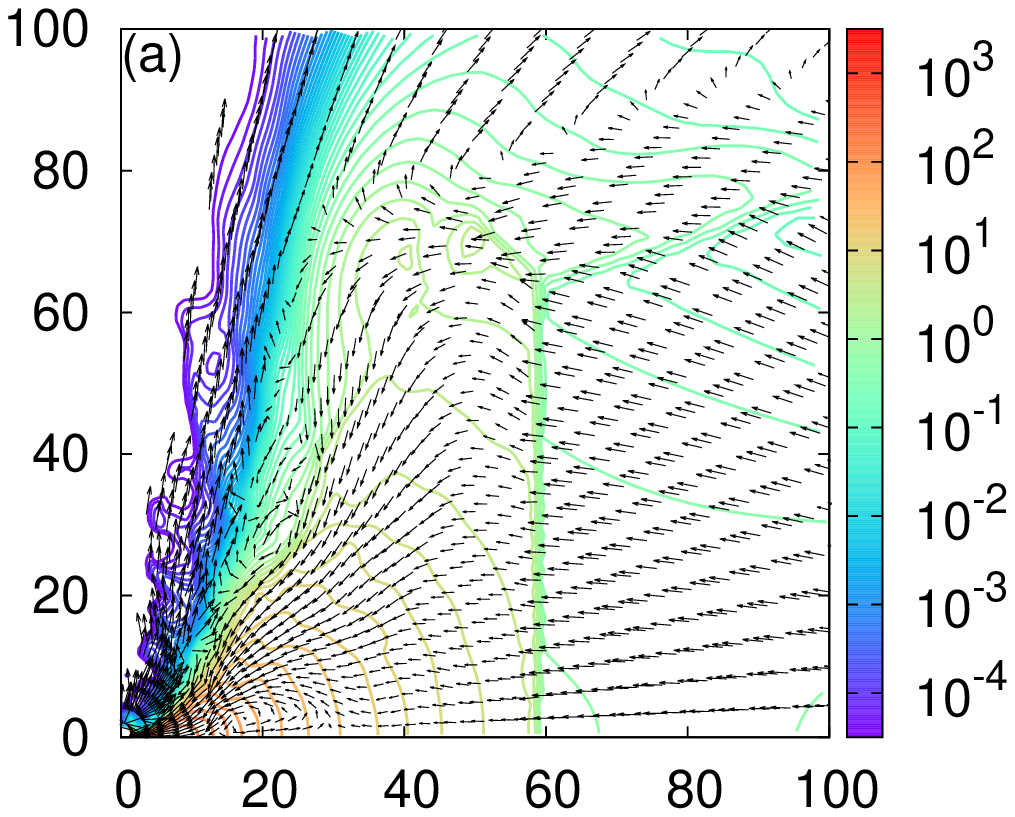}
\includegraphics[width=\columnwidth]{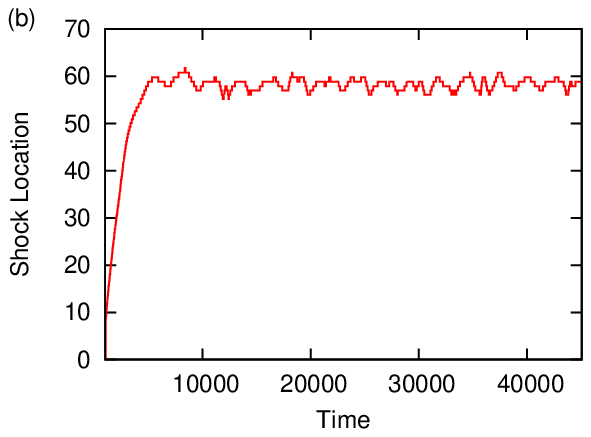}
\caption{Analogous figure to \autoref{fig:SA} for the retrograde case ($a=-0.95$).
We choose parameters corresponding to run R2 for this case.
Analytically, for these parameters, a steady shock is found to be formed at $25.8$.
However, in numerical simulation, a shock is found to be formed at $\sim 58$.}
\label{fig:SA_retro}
\end{figure}

In \autoref{fig:SA_retro}(a)-(b) we show analogous results to \autoref{fig:SA} 
when the black hole spinning in the opposite direction as that of the accretion 
(run R2). We choose $a=-0.95$ in this case. We choose, $\epsilon=1.001$ 
and $l=4.0$ which gives $V=0.0609$ and $a_s=0.0363$ at $r_{\rm out}=200$
for a model flow in vertical equilibrium. We use this velocity to inject matter at the outer boundary. Theoretical location of 
the shock for these input parameters is $25.8$. However, presence of turbulence pushes the shock to about $58$. There is also 
a small amplitude oscillation. Because of the effects of dragging of the inertial frame close to the black hole, forcing
the matter to co-rotate with it, there is a large turbulent cell in the CENBOL region. Also note that the oblique and triple shock 
being weaker in this case, the flow tends to be deflected to form a stronger wind. It is thus possible that 
the contra-rotating black holes would produce more profuse winds and this phenomenon certainly deserves attention. 

\begin{figure}
\includegraphics[width=\columnwidth]{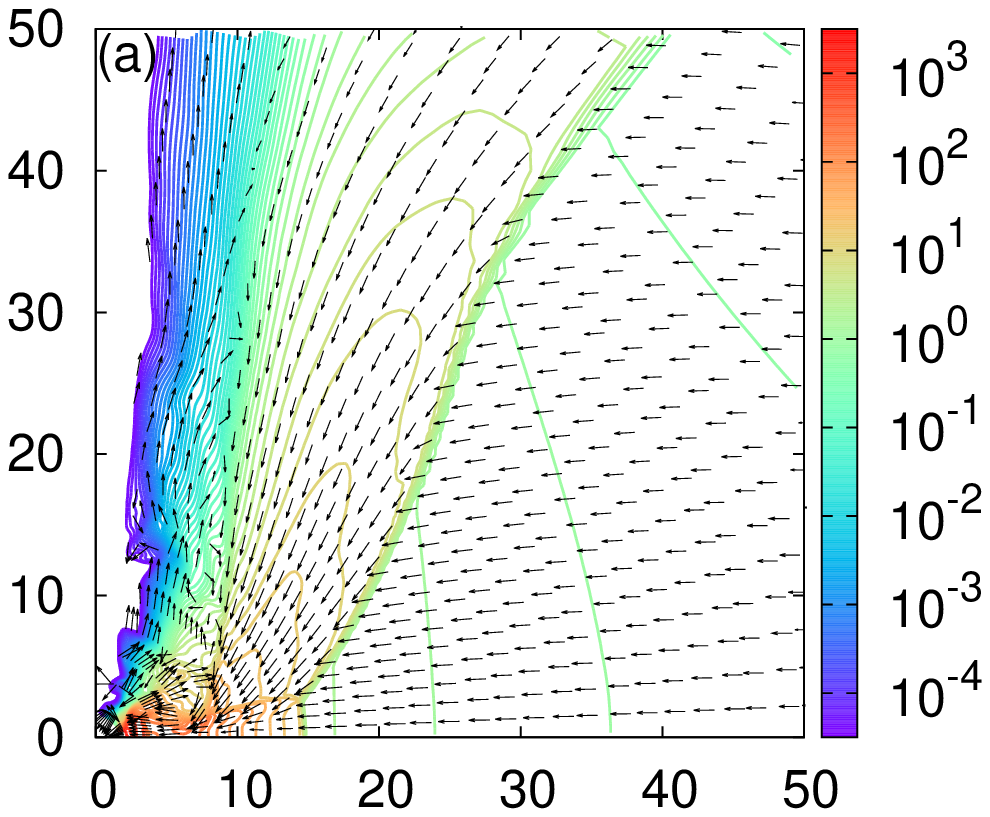}
\includegraphics[width=\columnwidth]{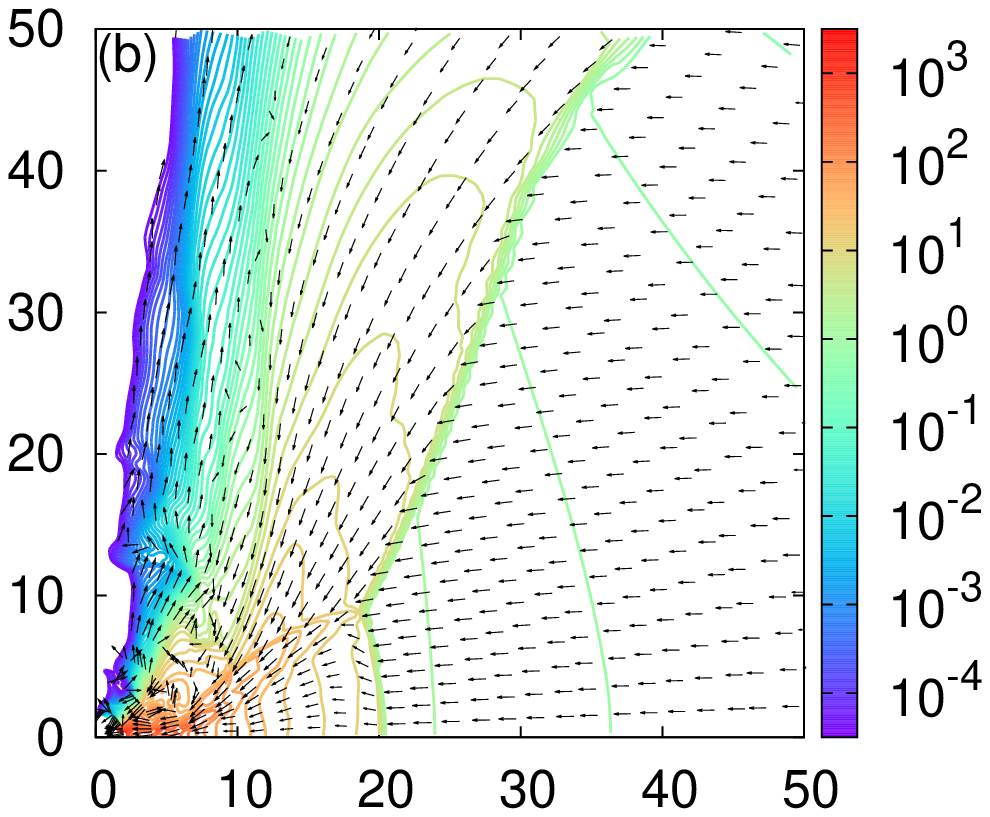}
\includegraphics[width=\columnwidth]{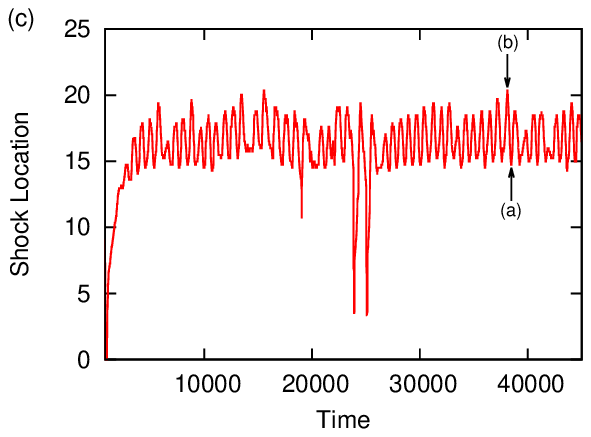}
\caption{Results of our simulation when the parameters $\epsilon$ and $l$ are not supposed to have 
shocks in a flow in vertical equilibrium (run R3). (a)-(b): Snapshots of density contours over-plotted with velocity vectors
at two different times $38500$ and $38100$, respectively. The time variation of shock location
on the equatorial plane is shown in (c) which show very large amplitude shock oscillations.
(a) and (b) are drawn at the times marked by the arrows on (c).}
\label{fig:NSA}
\end{figure}

Two dimensional simulations reveal many new aspects of the physics of accretion flows. For instance,
the presence of turbulence may change the topology of the solutions altogether. We show example of one such 
case (run R3) in the next simulation, where we inject a flow which is not supposed to have any standing shock 
in the vertical equilibrium model (i.e., from the region `NSA' of \citet{Chakrabarti1996c}). 
Here we choose $a=0.95$ and inject with flow parameters of $\epsilon=1.001$ and $l=2.21$ 
which gives $V=0.0634$ and $a_s=0.0362$ at $r_{\rm out}=200$. The flow is injected with this 
velocity as before. We observe that a shock forms nevertheless, but it does not remain steady. Rather, it exhibits a large amplitude oscillations. 

In \autoref{fig:NSA}(a)-(b), we plot the density contours superposed with velocity vectors. In \autoref{fig:NSA}(a), the shock is located at $\sim 15$
while in \autoref{fig:NSA}(b), the shock is located at $\sim 20$. In \autoref{fig:NSA}(c), instantaneous shock locations 
are plotted. This oscillation of shock is very important from the observational point of view, since
these are believed to be responsible for Quasi-Periodic Oscillations of X-rays amplitudes 
from black hole accretion disks.

\begin{figure}
\includegraphics[width=\columnwidth]{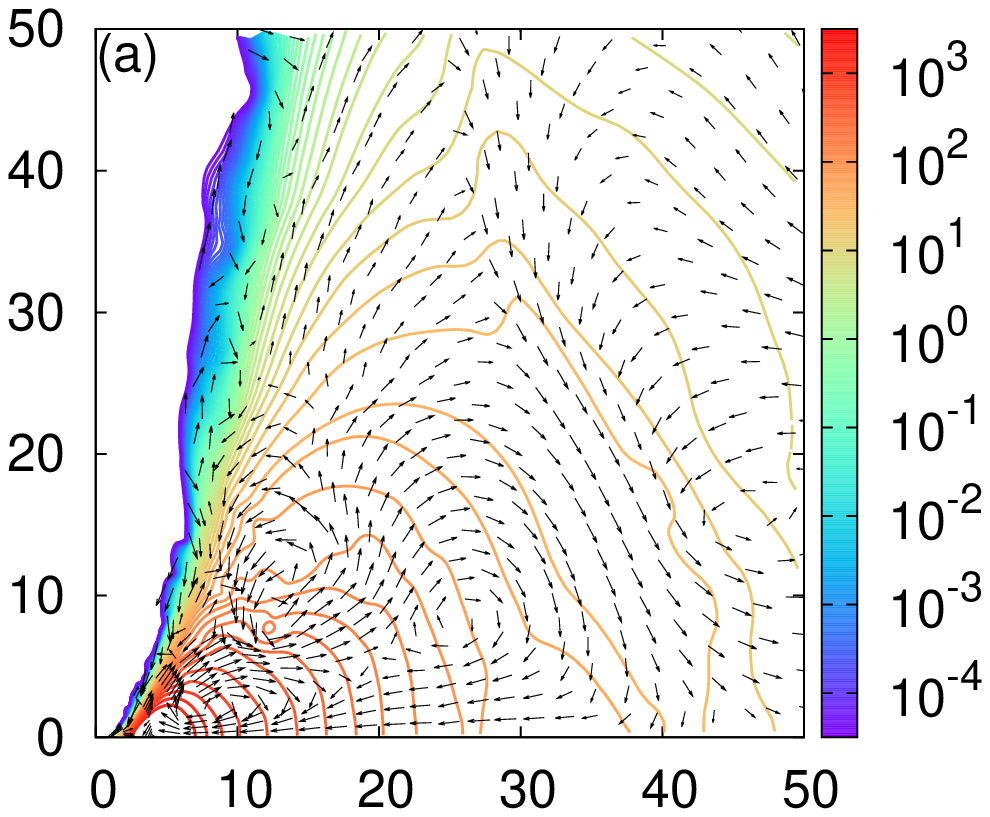}
\includegraphics[width=\columnwidth]{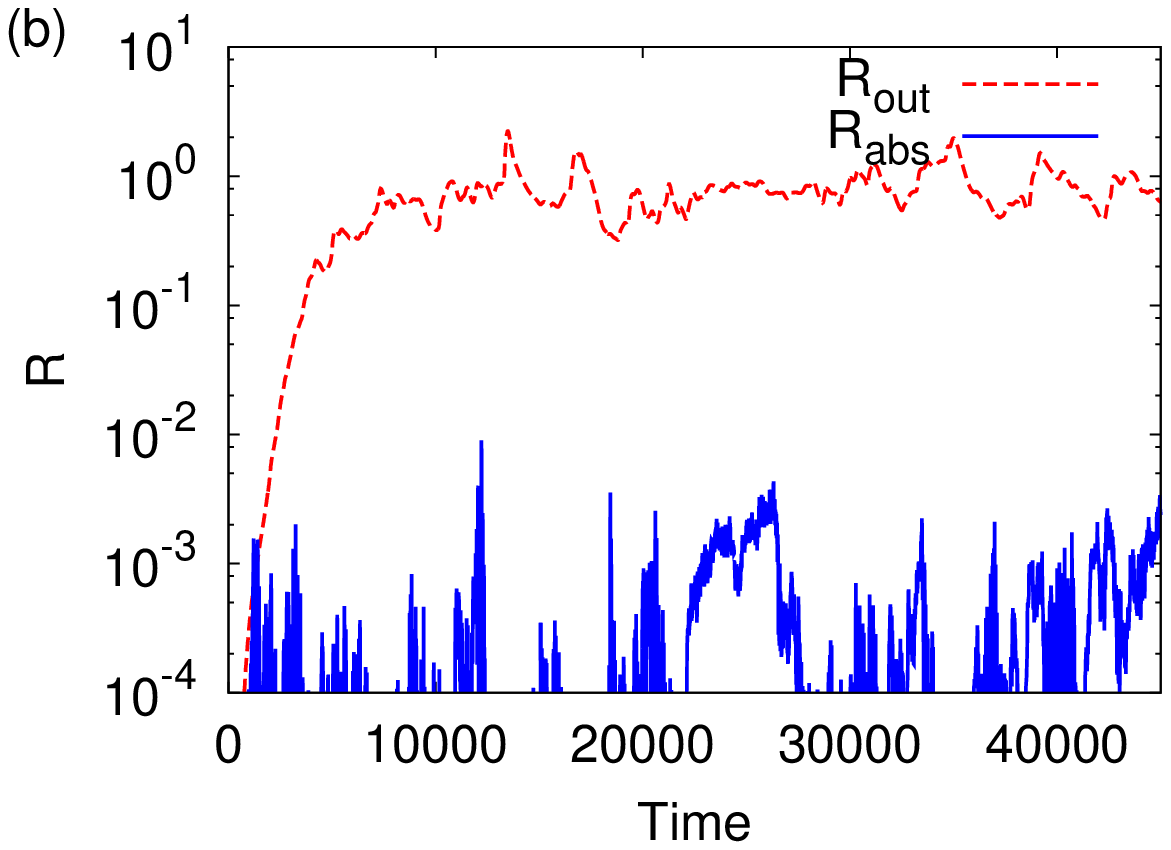}
\caption{Results of a case with the parameters chosen from O$^*$ region (run R4).
(a)Density contours over-plotted with velocity vectors showing a thick accretion disk pattern in the absence of accretion.
(b) Ratio of the mass absorption rate to the mass inflow rate 
($R_{\rm abs}=\dot{M}_{\rm abs}/\dot{M}_{\rm in}$) is shown in blue and 
Ratio of the mass outflow rate to the mass inflow rate 
($R_{\rm out}=\dot{M}_{\rm out}/\dot{M}_{\rm in}$)is shown in red.
Nearly no matter is absorbed by the black hole and almost all the matter is leaving through the outer boundary of the 
computational domain. Occasional $R_{\rm out} > 1$ indicates that the disk is evacuated at that time.}
\label{fig:LMB}
\end{figure}

In \autoref{fig:LMB}(a)-(b) we show results of another case (run R4), where the flow parameter is
from `O$^{*}$' region \citep{Chakrabarti1996c}. The flow parameters are chosen as $\epsilon=1.001$ and $l=2.63$ which yields
$V=0.06299$ and $a_s=0.0362$ at $r_{\rm out}=200$ from the transonic solution. 
In this case, since the specific angular momentum is larger than the marginally bound value, 
the steady flow in vertical equilibrium model does not even extend till the horizon and thus a steady flow is impossible. 
When a time dependent simulation is 
run we see a very interesting behavior. The flow is found to be full of turbulence in large and smaller scales, as the branch to enter
the black hole is practically blocked due to high centrifugal barrier. Indeed without accretion, and mostly rotating, 
the isodensity contours assume the shape of thick accretion disks \citep{Paczynsky1980a,Chakrabarti1985}. 
The flow tries to enter into the black hole but is unable to do so resulting in accretion of a very little matter.
From the simulation, we compute the mass absorption rate, $\dot{M}_{\rm abs}$,
at which the matter is absorbed by black hole through the inner boundary of the computational domain and
the mass outflow rate, $\dot{M}_{\rm out}$, at which the matter leaves the computational
domain through the outer boundary as outflow or jet.
$\dot{M}_{\rm in}$ measures the mass inflow rate of the incoming matter through the in-flowing boundary. 

In \autoref{fig:LMB}(a), we show the contours of constant density superposed with velocity vectors. 
In \autoref{fig:LMB}(b), the blue curve shows the ratio of the mass absorption rate to the mass inflow rate
($R_{\rm abs}=\dot{M}_{\rm abs}/\dot{M}_{\rm in}$).
The red curve shows the ratio of the mass outflow rate to the mass inflow rate
($R_{\rm out}=\dot{M}_{\rm out}/\dot{M}_{\rm in}$).
As we can see, nearly no matter is absorbed by the black hole for this case.
Instead, almost all the matter is leaving through the outer boundary of the
computational domain (the ratio is around unity for red line).
Sometimes, this ratio is more than one
showing that the accumulates disk matter may also be evacuated in this case. Figure 6b is drawn 
in logarithmic scale in order to show small and sudden inflows into the black hole, possibly
because of transport of angular momentum of some matter to those leaving the system. Such transports
are common in turbulent disks through turbulent viscosity.

For all the cases presented in \autoref{table:1}, we can estimate Compton y-parameter 
\citep{Rybicki1979} in the pre- and post-shock regions. Assuming a mass accretion rate $\sim$ 
0.1 Eddington rate around a 10$M_\odot$ black hole, the values of the Compton y-parameter are
found to be $y_{\rm preshock} \sim 0.0016 $ and $y_{\rm postshock} \sim 10$. Precisely, such 
consideration is used to study spectral properties by the TCAF solution \citep{Chakrabarti1995a}.

\begin{figure*}
\includegraphics[width=\columnwidth]{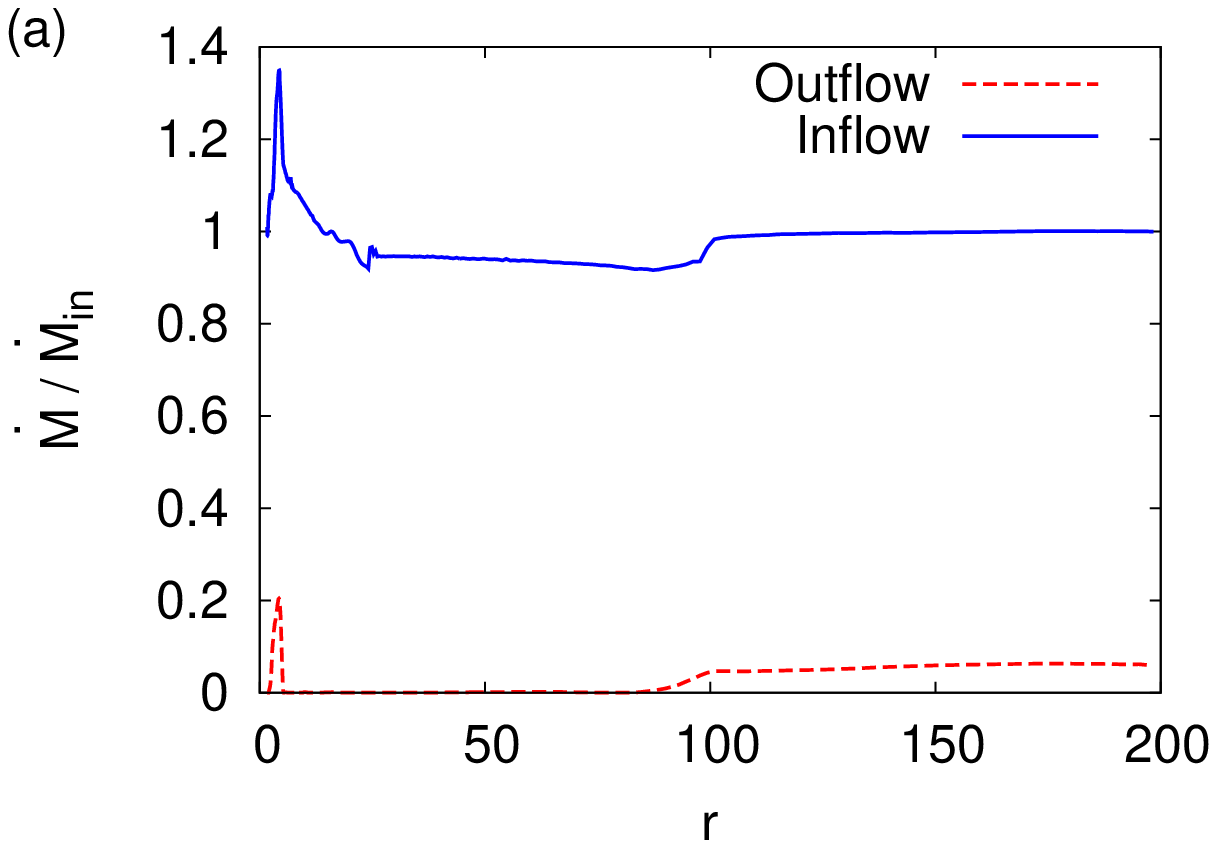}
\includegraphics[width=\columnwidth]{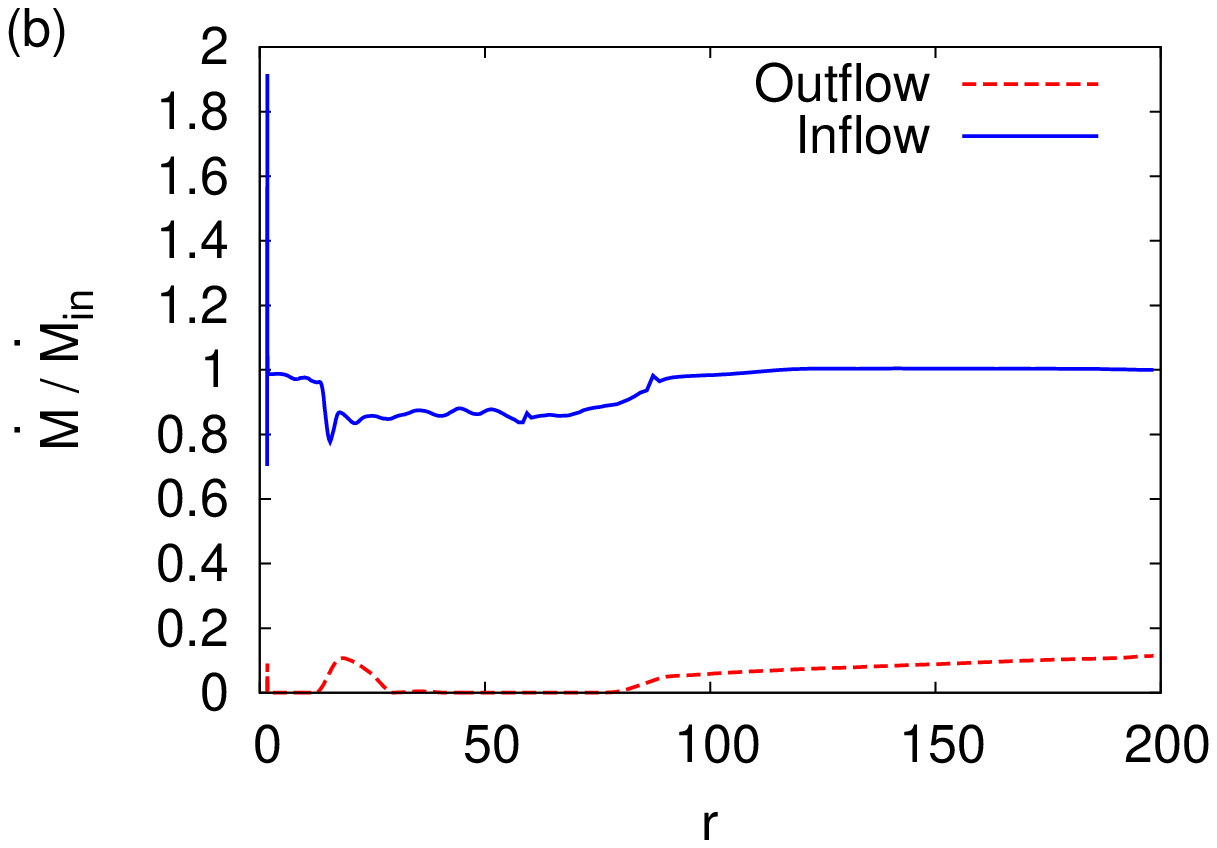}
\includegraphics[width=\columnwidth]{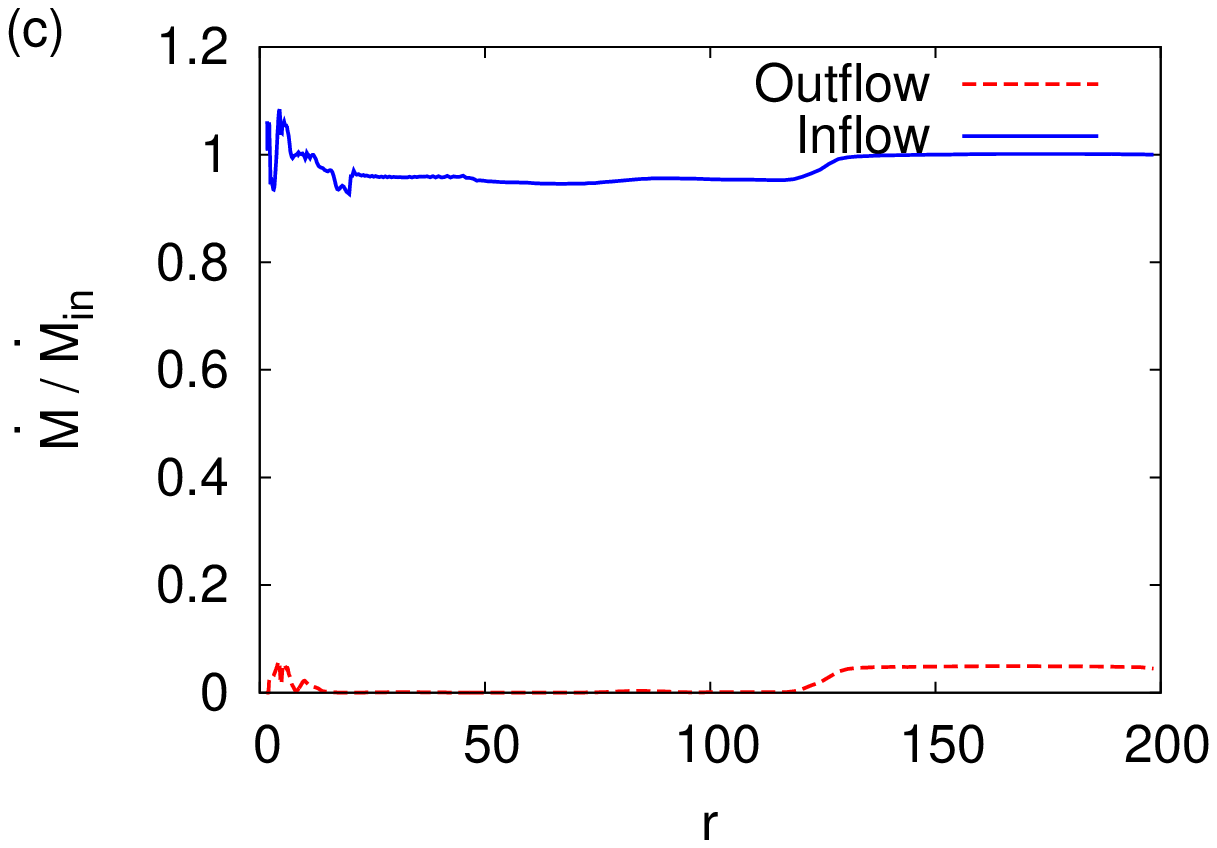}
\includegraphics[width=\columnwidth]{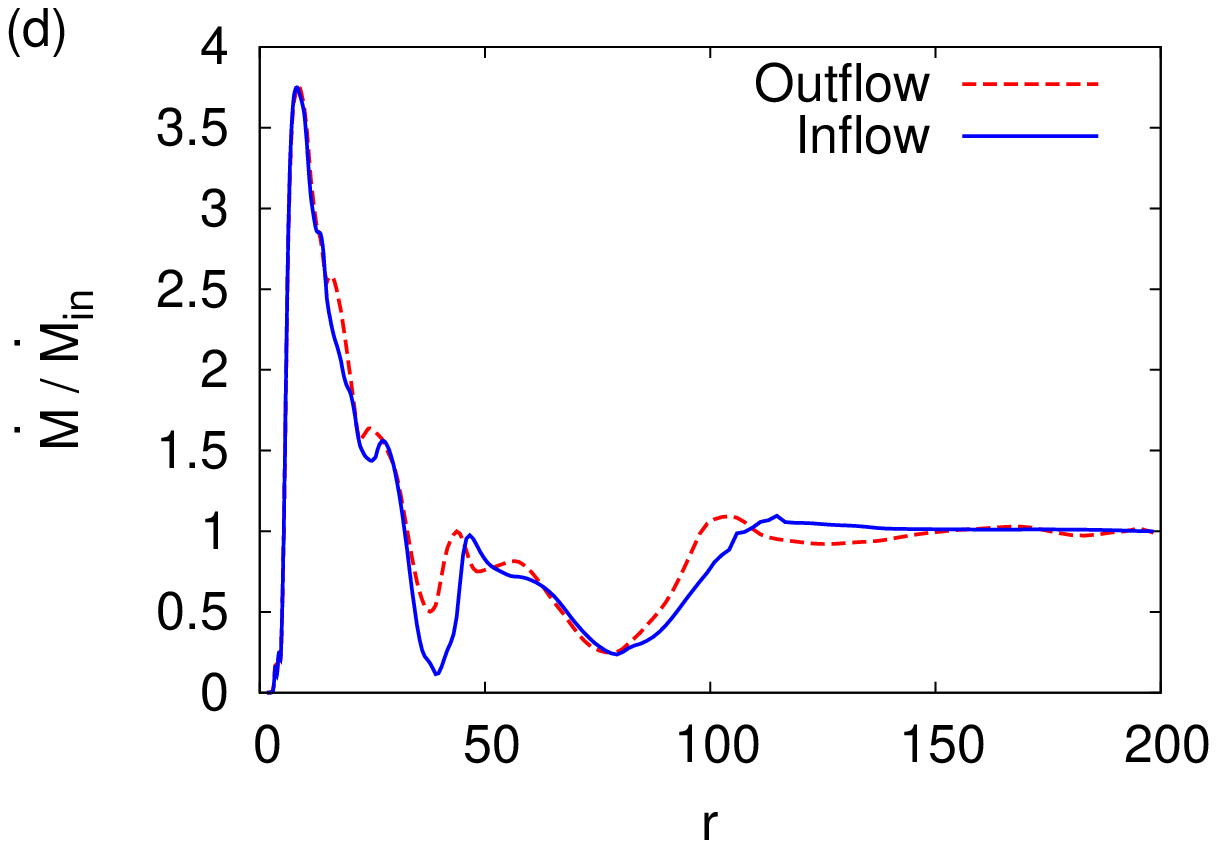}
\caption{Inflow-Outflow rates, normalized by the mass accretion rate $dot{M}_{\rm in}$ at $r_{\rm out}$,
as a function of radius for all the cases presented in \autoref{table:1}. Blue, solid
curve shows the inflow rate and red, dashed curve shows the outflow rate. (a), (b)
(c) and (d) correspond to cases R1, R2, R3 and R4, respectively.}
\label{fig:flux}
\end{figure*}

In our two-dimensional simulations, we also notice the formation
of the outflow from the surface of the CENBOL. In order to understand what
fraction of injected matter can be swallowed by black hole and what fraction 
can leave as outflow, we plotted the radial variation of the corresponding rates
for all the four runs presented in \autoref{table:1}. \autoref{fig:flux} shows the 
results in the steady state for all the cases. 
In order to find the total inflow rate at a particular radial point, 
we integrated the inward pointing mass flux in the angular direction at that 
radius. Similarly, to find the total outflow rate at a particular radial point, 
we integrated the outward pointing mass flux in the angular direction at that  
radius. From \autoref{fig:flux}(a), (b) and (c), we find that the flow is mostly accreting
at all radii and a small fraction ($\sim$ 5\%) of injected matter points away
from the black hole. \autoref{fig:flux}(d), on the other hand, shows that
at all radii, the inward and the outward mass rate are roughly equal.

Please note that in this paper, our primary goal is to verify whether the numerical solutions are
consistent with the theoretical solutions provided in \citet{Chakrabarti1996b,Chakrabarti1996c}.
During the process, we also find from our simulations that outflow can originate from the
post-shock region. Although, our current simulations show that the outflow rate
is not significant for the parameters that we use, the rate can be enhanced for accreting flow
with higher $l$ \citep[see, e.g.,][]{Molteni1994a}. 

\section{Conclusions}
\label{sec:sec4}

Accretion processes on a black hole continues to remain a fascinating subject and with the advent of better observational tools it has become possible to prove models of accretion flows. It is widely understood that an astrophysical black hole may not have any significant Keplerian disk 
in most of the time and the accretion may be wind driven, i.e., have low angular momentum.
This is especially true for high mass X-ray binaries and Active Galactic Nuclei. It is also found that most of the black holes have spins \citep[see review by][]{Remillard2014}. So it is essential to carry out the studies in full Kerr geometry. In the literature, systematic studies have been presented \citep{Chakrabarti1996b,Chakrabarti1996c} when the low angular momentum flow is steady and geometrically thin modeled as a conical wedge flow or a flow in vertical equilibrium. However, no theoretical study is either present or possible in Kerr geometry for a geometrically unrestricted flow. 

In this paper, for the first time, we present results of a fully general relativistic flow. We first tested the code in one dimension (i.e. on the equatorial plane) and showed that the code exactly reproduced the corresponding flow solutions with and without a shock transition, in both accretion and winds. We then applied the code for a two dimensional Bondi flow and found that due to dragging of the inertial frame, even a spherically symmetrically injected flow at the outer grid boundary takes an oblate shape and the isodensity contours show this deformity. Finally, we ran several cases of two dimensional flows with finite specific angular momentum. We chose input parameters and compared the results with steady solutions from vertical equilibrium model. We discover that turbulent pressure is much stronger for a rapidly spinning black hole and pushes the shock by a large amount. This effect was much weaker in Schwarzschild black hole modeled with pseudo-Newtonian potential \citep{Molteni1994a}. We also carried out simulations with contra-rotating black holes and show that the flow is very turbulent close to the horizon, especially because of the dragging of matter in the opposite direction by the black hole. We also showed that because of this turbulence,
the shock is located much further out. Finally, we considered a case which is not supposed to be accreting according to vertical equilibrium model.
Indeed, we find that the accretion is very very low and almost all the matter is ejected as winds, often partially evacuating the disk itself when the
outflow rate is higher than the inflow rate.
In a realistic and generalized accretion process scenario on a black hole, one often finds signatures of two components where the viscous 
accretion disks are surrounded by low-angular momentum flows such as those we used in this paper 
\citep{Chakrabarti1996a,Giri2013a,Giri2015a,Chakrabarti2015a}.
In future, our goal would be to introduce viscosity and radiative processes to produce self-consistent spectrum 
out of disks around a rotating black hole. In that case the spin parameter may also be determined from the value of frequency at which the shocks oscillate. These would be carried out in future.

\section*{Acknowledgements}
DSB acknowledges support via NSF grants NSF-ACI-1533850, NSF-DMS-1622457, NSF-ACI-1713765.
Several simulations were performed on a cluster at UND that is run by the Center for Research Computing.
Computer support on NSF's XSEDE and Blue Waters computing resources is also acknowledged.
JK also acknowledges support by the National Research Foundation of Korea grant NRF-2018R1D1A1B07042949.




\bibliographystyle{mnras}
\bibliography{references} 





\bsp	
\label{lastpage}
\end{document}